\documentclass[aps,pra,floatfix,superscriptaddress,twocolumn]{revtex4-1}
\usepackage[dvips]{graphicx}
\usepackage{color}

\usepackage{longtable}
\usepackage{dcolumn}
\usepackage[dvips]{graphicx}
\usepackage{bm}
\usepackage{bbm}

\usepackage{times}
\usepackage{nicefrac}
\usepackage{amsmath}
\usepackage{amsfonts}
\usepackage{amssymb}
\usepackage{amsthm}

\newcolumntype{w}[1]{D{.}{.}{#1}}

\newcolumntype{.}{D{x}{}{-1}}

\newcommand{\balpha}{\bm{\alpha}}

\newcommand{\bfr}{{\bm {r}}}

\newcommand{\bfp}{{\bm {p}}}

\newcommand{\Za}{Z\alpha}

\begin{document}

\title{Non-linear isotope-shift effects in Be-like, B-like, and C-like argon}

\author{V.~A. Yerokhin}
\affiliation{Physikalisch-Technische Bundesanstalt, D-38116 Braunschweig, Germany}
\affiliation{Center for Advanced Studies, Peter the Great St.~Petersburg Polytechnic University,
195251 St.~Petersburg, Russia}

\author{R.~A.~M\"uller}
\affiliation{Physikalisch-Technische Bundesanstalt, D-38116 Braunschweig, Germany}
\affiliation{Technische Universit\"at Braunschweig, D-38106 Braunschweig, Germany}

\author{A.~Surzhykov}
\affiliation{Physikalisch-Technische Bundesanstalt, D-38116 Braunschweig, Germany}
\affiliation{Technische Universit\"at Braunschweig, D-38106 Braunschweig, Germany}

\author{P.~Micke}
\affiliation{Physikalisch-Technische Bundesanstalt, D-38116 Braunschweig, Germany}
\affiliation{Max-Planck-Institut f\"ur Kernphysik, D-69117 Heidelberg, Germany}

\author{P.~O.~Schmidt}
\affiliation{Physikalisch-Technische Bundesanstalt, D-38116 Braunschweig, Germany}
\affiliation{Institut f\"ur Quantenoptik, Leibniz Universit\"at Hannover, 30167 Hannover, Germany}

\begin{abstract}

Violation of linearity of the King plot is investigated for a chain of partially stripped argon
isotopes. The nonlinearity originates within the Standard Model from subtle contributions to the
isotope shifts from next-to-leading order effects, which have never been systematically studied
so far. In light atoms these nonlinear effects are dominated by the quadratic nuclear recoil
($\propto 1/M^2$ where $M$ is the nuclear mass). Large-scale relativistic calculations of the
linear and quadratic mass shift and the field shift are performed for the $2P$ fine-structure
transitions in Be-like, B-like, and C-like argon ions. Nonlinearities of the King plots from 5 to
30~kHz are found, which is four orders of magnitude larger than previous estimates in comparable
systems. Accurate calculations of these effects are vital for identification of possible
nonlinearities originating from physics beyond the Standard Model.

\end{abstract}

\maketitle

\section{Introduction}

Investigations of isotope-shift phenomena offer an excellent possibility to selectively probe
nuclear effects and to extract nuclear parameters from the observed atomic spectra. On the
theoretical side, the isotope shifts of atomic levels have the advantage that they can be
calculated to a much higher absolute precision than the atomic energy levels. Depending on the
nuclear charge $Z$, the isotope shifts are governed either by the nuclear mass (low-$Z$ ions) or by
the finite nuclear size (high-$Z$ ions), thus yielding an opportunity for a detailed study of
individual nuclear effects. On the experimental side, the isotope-shift phenomena offer a
possibility for extracting information about the nucleus, by means of the so-called King-plot
analysis \cite{king:63,king:84}.

Isotope-shift studies can also improve our understanding of fundamental physics. It has recently
been demonstrated \cite{frigiuele:17,berengut:18} that isotope-shift measurements can be used to
constrain the coupling strength of hypothetical new-physics boson fields to electrons and neutrons.
More specifically, the presence of a light boson particle would cause a nonlinearity of the King
plot for the isotope shifts of two atomic transitions of several isotopes of the same element. The
absence of the King-plot nonlinearities observed so far allowed the authors to draw constraints on
the coupling strength of the hypothetical particles.

Experimentally, no King-plot nonlinearities were observed in the measurements
\cite{gebert:15,shi:16} performed at the 100~kHz accuracy level. The present-day isotope-shift
experiments, however, may improve the accuracy by several orders of magnitude. Specifically,
measurements of optical-clock transitions were demonstrated on a few-Hertz precision level, by
simultaneously exciting two Ca$^+$ isotopes in the same trap \cite{knollmann:19}. An even higher
precision can be achieved by using correlated or even entangled states \cite{manovitz:19}. The
coherent high-resolution optical spectroscopy \cite{micke:19:priv} can provide access to
isotope-shift measurements of highly-charged ions, thus tremendously extending the choice of useful
transitions.

As already pointed out in Ref.~\cite{berengut:18}, some small nonlinearities of the King plot
should appear within the Standard Model framework, but they have never been calculated so far. The
only attempt to address this issue was made by Flambaum and co-workers~\cite{flambaum:18}, who
derived approximate analytical formulas for the field shift in the mean-field approximation and
calculated the King-plot nonlinearities for heavy and superheavy atoms. In the present work we
perform relativistic calculations of the nonlinear isotope-shift effects for several fine-structure
transitions in argon ions. We also analyse constraints on hypothetical boson fields that can be
realistically derived from the King-plot analysis in these systems.

The isotope-dependent part of the energy of an electronic state of an atom is traditionally
represented as a sum of the {\em mass shift} and the {\em field shift},
\begin{eqnarray}\label{eq:1}
E_{\rm is} = \frac{m}{M}\, K +  \frac{R^2}{\lambdabar_C^2}\,F\,,
\end{eqnarray}
where $m$ is the electron mass, $M$ is the nuclear mass, $R = \big< r^2 \big>^{1/2}$ is the
root-mean-square (rms) radius of the nuclear charge distribution, and $\lambdabar_C$ is the Compton
wavelength divided by $ 2\pi$ ($\lambdabar_C = 386.159$~fm, $ \lambdabar_C  = 1$ in relativistic
units). $K$ and $F$ are usually called the mass-shift and the field-shift constant, respectively.
Note that in our formulation, the constants $K$ and $F$ have units of energy, since they are
multiplied by dimensionless ratios in Eq.~(\ref{eq:1}).

It is important that in the present work we require Eq.~(\ref{eq:1}) to be {\em exact} in the
Standard Model framework. In other words, we ascribe all higher-order effects to $K$ and/or to $F$,
which thus acquire some dependence on nuclear parameters. In many practical situations, this weak
dependence can be ignored and one can treat $K$ and $F$ as ``constants'' depending only on the
electronic state of the atom but not on the nuclear properties of the isotope. As explained below,
such an assumption leads to a King plot which is exactly linear. In the present work, however, we
will address deviations from this linear form, caused by a tiny dependence of $K$ and $F$ on
nuclear parameters.

\section{King plot}

The King plot \cite{king:63,king:84} is a widely used method that allows for a systematic study of
the isotope shifts of two atomic transitions in a chain of isotopes. In order to construct a King
plot, we consider two electronic transitions (which will be labelled as ``$a$'' and ``$b$'') for a
chain of at least four isotopes of the same element with mass numbers $(A_0,A_1,A_2,\ldots)$. Note
that the transitions $a$ and $b$ may belong to different charge states of the same element.

\subsection{Standard formulation}

Within the standard formulation, the mass-shift and field-shift constants in Eq.~(\ref{eq:1}) are
assumed to depend only on the electronic transition but not on the isotope. In this case, the
isotope shift of the energy of the transition $a$ between the isotopes $i$ and $j$ is
\begin{align}\label{eq:2}
E_{aij} &\  = \Big(\frac{m}{M_i}-\frac{m}{M_j}\Big)\, K_a +
 \bigg(\frac{R^2_i}{\lambdabar_C^2}-\frac{R^2_j}{\lambdabar_C^2}\bigg)\,F_a\,,
 \nonumber \\ & \
  \equiv {\cal M}_{ij}\,K_a + {\cal R}_{ij}\,F_a\,.
\end{align}
Introducing the modified transition energies $n_{aij}$,
\begin{align}\label{eq:3}
n_{aij} &\  = \frac{E_{aij}}{{\cal M}_{ij}}\,,
\end{align}
one rewrites Eq.~(\ref{eq:2}) as
\begin{align}\label{eq:4}
n_{aij} &\  = K_a + \frac{{\cal R}_{ij}}{{\cal M}_{ij}}\,F_a\,.
\end{align}
Considering Eq.~(\ref{eq:4}) for two transitions $a$ and $b$, one can eliminate the
isotope-dependent constant ${\cal R}_{ij}/{\cal M}_{ij}$, arriving at
\begin{align}\label{eq:5}
n_{bij} &\  = \Big(K_b - \frac{F_b}{F_a} \, K_a\Big) + \frac{F_b}{F_a}\,n_{aij}\,.
\end{align}
Fixing the index $j$ and plotting $n_{bij}$ ($=y_i$) against $n_{aij}$ ($=x_i$) for different
isotopes $i$, one gets the linear dependence of the form
\begin{align}\label{eq:6}
y_i = {\cal A} + {\cal B}\,x_i\,,
\end{align}
where the coefficients ${\cal A}$ and ${\cal B}$ do not depend on the isotope parameters. The
dependence of $n_{bij}$ on $n_{aij}$ is widely known as the King plot \cite{king:63,king:84}.
Measuring the modified frequencies $n_{bij}$ and $n_{aij}$, one obtains the experimental values for
the coefficients ${\cal A}$ and ${\cal B}$, {\em i.e.}, for the ratio of the field-shift constants
${F_b}/{F_a}$ and for the combination $K_b - ({F_b}/{F_a}) \, K_a$.

\subsection{Extended formulation: standard model}

We now take into account that the constants $K$ and $F$ in Eq.~(\ref{eq:1}) depend not only on the
transition but also on the isotope, $K \equiv K_{ai} = K_a + \delta K_{ai}$ and the same for $F$.
In this case, Eq.~(\ref{eq:4}) becomes
\begin{align}\label{eq:7}
n_{aij} &\  = K_{aij} + \frac{{\cal R}_{ij}}{{\cal M}_{ij}}\,F_{aij}\,,
\end{align}
where
\begin{align}\
K_{aij} &\ = \frac{\frac{m}{M_i}\, K_{ai}-\frac{m}{M_j}\, K_{aj}}{{\cal M}_{ij}} \equiv K_a + \delta K_{aij}\,,
 \\
F_{aij} &\ = \frac{\frac{R_i^2}{\lambdabar_C^2}\,F_{ai} - \frac{R_j^2}{\lambdabar_C^2}\,F_{aj}}{{\cal R}_{ij}}
 \equiv F_a + \delta F_{aij}\,.
\end{align}
Eq.~(\ref{eq:5}) then becomes
\begin{align}\label{eq:9}
n_{bij} &\  = \Big(K_{bij} - \frac{F_{bij}}{F_{aij}} \, K_{aij}\Big) + \frac{F_{bij}}{F_{aij}}\,n_{aij}\,.
\end{align}

Considering the above equation as a functional dependence of $n_{bij}$ ($=y_i$) on $n_{aij}$
($=x_i$) for different values of the isotope index $i$ and a fixed $j=0$, we get a set of equations
\begin{align}\label{eq:10}
y_i = {\cal A}_i + {\cal B}_i\,x_i\,.
\end{align}
The coefficients ${\cal A}_i$ and ${\cal B}_i$ in the above equations depend (slightly) on the
isotope index $i$ and, therefore, the (three or more) points $(y_i,x_i)$ nor longer lie on a
straight line.

Restricting to the minimal number of three points, the nonlinearity of a 3-point curve
(\ref{eq:10}) may be conveniently defined \cite{flambaum:18} as a shift of the ordinate of the
third point from the straight line defined by the first two points,
\begin{align}\label{eq:11}
\delta y = \big( y_3-y_1\big) - \frac{y_2-y_1}{x_2-x_1}\,\big( x_3-x_1\big)\,.
\end{align}
Rewriting this definition for the King plot, we arrive at
\begin{align}\label{eq:12}
\delta E_{b30} &\  = {\cal M}_{30}\bigg[
  n_{b30}-n_{b10}
\nonumber \\ &
  - \frac{n_{b20}-n_{b10}}{n_{a20}-n_{a10}}\,\big( n_{a30}-n_{a10}\big)
  \bigg] \,.
\end{align}
Note that $\delta E_{b30}$ has the unit of energy. Physically, it is the difference of the
$(A_3,A_0)$ isotope shift of the transition $b$ from the linearly-predicted position based on the
$(A_2,A_0)$ and $(A_1,A_0)$ isotope shifts of the transitions $a$ and $b$.

$\delta E_{b30}$ is the definition of the King-plot nonlinearity as used in
Ref.~\cite{flambaum:18}. It has, however, a drawback of being not symmetrical with respect to the
transitions $a$ and $b$. In other words, with just three points, there are two different
nonlinearities, $\delta E_{b30}$ and $\delta E_{a30}$ (where $\delta E_{a30}$ is obtained from
$\delta E_{b30}$ by $a\leftrightarrow b$). In the present work, we define the nonlinearity in a
symmetric way, as a half-sum of these two absolute values,
\begin{align}\label{eq:12b}
\Delta_{\rm NL}(ab) = \Delta_{\rm NL}(ba) = \frac12 \big( \big| \delta E_{a30}\big| +
\big| \delta E_{b30}\big|\big)\,.
\end{align}

\subsection{Extended formulation: new physics}

We now consider the King-plot analysis in the presence of a hypothetical  boson particle with mass
$m_{\phi}$. The interaction between the electrons and neutrons mediated by such a boson
 can be effectively described \cite{delaunay:17} by a Yukawa-type potential ,
\begin{eqnarray}\label{eq:np:1}
V_{\phi} = -\alpha_{\rm NP}\,\big(A-Z\big)\,\frac{e^{-m_{\phi}\,r}}{r}\,,
\end{eqnarray}
where $A-Z$ is the number of neutrons in the nucleus and $\alpha_{\rm NP}$ is the coupling
constant, $\alpha_{\rm NP} = q_nq_e$, where $q_n$ and $q_e$ are the strength of coupling to
neutrons and electrons, respectively.

With a new particle, the isotope-dependent part of the energy of the reference state becomes ({\em
cf.} Eq.~(\ref{eq:1}))
\begin{eqnarray}\label{eq:1b}
E_{\rm is} = \frac{m}{M}\, K +  \frac{R^2}{\lambdabar_C^2}\,F  + \frac{\alpha_{\rm NP}}{\alpha}\,A\,X_{\phi}\,,
\end{eqnarray}
where $A$ is the mass number of the isotope and $\alpha$ is the fine-structure constant.
$X_{\phi}$ is the ``new-physics'' isotope-shift constant defined as
\begin{eqnarray} \label{eq:7c}
X_{\phi} = \Big< - \sum_k \frac{\alpha \, e^{-m_{\phi}\,r_k}}{r_k}\Big>\,,
\end{eqnarray}
where $k$ numerates the electrons in the atom and the matrix element is evaluated with the atomic
reference-state wave function. The expression for the reduced frequency now becomes ({\em cf.}
Eq.~(\ref{eq:7}))
\begin{align}\label{eq:7b}
n_{aij} &\  = K_{aij} + \frac{{\cal R}_{ij}}{{\cal M}_{ij}}\,F_{aij}
  + \frac{\alpha_{\rm NP}}{\alpha}\,\frac{A_i-A_j}{{\cal M}_{ij}}\,X_{\phi, a}  \,,
\end{align}
where we took into account that the isotope dependence of $X_{\phi, a}$ can be safely neglected.

\section{Theory of the isotope shift}

\subsection{Leading effects}

The mass shift of energy levels is induced by the nuclear recoil effect. Within the Breit
approximation ({\em i.e.}, up to the order $(Z\alpha)^4 m/M$), the recoil effect is induced by the
relativistic recoil operator \cite{shabaev:85,shabaev:98:rectheo}
\begin{align}\label{eq:rrec}
H_{\rm rec} \equiv \frac{m}{M}\,\widetilde{H}_{\rm rec} =  \frac{m}{M}\,\Big( \widetilde{H}_{\rm rnms}
 + \widetilde{H}_{\rm rsms}\Big)\,,
\end{align}
where ${H}_{\rm rnms}$ and ${H}_{\rm rsms}$  are the relativistic normal and specific mass shift
operators, respectively,
\begin{eqnarray}
\widetilde{H}_{\rm rnms} = \frac12\,\sum_k \bigg[ \bfp_k^2
-\frac{Z\alpha}{r_k} \Big( \balpha_k + \frac{(\balpha_k\cdot\bfr_k)\,\bfr_k}{r_k^2} \Big)
\cdot\bfp_k \bigg]\,,
\nonumber\\
\end{eqnarray}
\begin{eqnarray}
\widetilde{H}_{\rm rsms} = \frac12\,\sum_{k\ne l} \bigg[ \bfp_k\cdot\bfp_l
-\frac{Z\alpha}{r_k} \Big( \balpha_k + \frac{(\balpha_k\cdot\bfr_k)\,\bfr_k}{r_k^2} \Big)
\cdot\bfp_l \bigg]\,,
\nonumber\\
\end{eqnarray}
and summations over $k$ and $l$ run over all electrons. Fully relativistic calculations of the
isotope-shift effects were performed  over the last two decades by several groups
\cite{tupitsyn:03,korol:07,zubova:14,naze:14,zubova:16}.

The leading, linear in $m/M$ mass-shift constant is given by the expectation value of the nuclear
recoil operator with the (non-recoil) atomic wave function of the reference state,
\begin{eqnarray} \label{eq:13a}
K^{(1)} = \big< \widetilde{H}_{\rm rec} \big>\,.
\end{eqnarray}

The field shift of energy levels is induced by the effect of the finite nuclear size (fns). The
leading field-shift constant can be obtained as an expectation value of the derivative of the
nuclear binding potential $V_{\rm nuc}$ over the square of the nuclear rms charge radius $R^2$
\cite{safronova:01}
\begin{align}
F = \big< V_{\rm FS} \big>\,,
\end{align}
where
\begin{align}
V_{\rm FS} = \sum_k \frac{\partial V_{\rm nuc}(r_k)}{\partial (R/\lambdabar_C)^2}\,,
\end{align}
and the summation over $k$ runs over all electrons.


\subsection{Quadratic mass shift}

The leading isotope-dependence of the mass-shift constant $K$ comes from the quadratic ($\propto
(m/M)^2$) nuclear recoil effect,
\begin{align}\label{eq:13}
K = K^{(1)} +  \frac{m}{M}\,K^{(2)}\,.
\end{align}
Within the Breit approximation, the quadratic mass shift is induced by the second-order
perturbation of the operator $H_{\rm rec}$. It is known \cite{pachucki:17:heSummary} that for a
spin-zero nucleus, there is no additional recoil operator $\propto (m/M)^2$ within the Breit
approximation.

We calculate the quadratic mass-shift constant $K^{(2)}$ in two steps. First, we construct the
nuclear-recoil-corrected many-electron wave function, by including the recoil operator
(\ref{eq:rrec}) into the Dirac-Coulomb-Breit Hamiltonian and diagonalizing the Hamiltonian matrix.
Second, we determine $K^{(2)}$, neglecting higher-order $\propto (m/M)^3$ effects, by taking the
difference
\begin{eqnarray} \label{eq:2ndrec}
\frac{m}{M}\,K^{(2)} = \frac12 \Big[ \big< \widetilde{H}_{\rm rec} \big>_M
   - \big< \widetilde{H}_{\rm rec} \big>\Big]\,,
\end{eqnarray}
where $\big<.\big>_M$ indicates the matrix element calculated with the nuclear-recoil-corrected
wave function and $\big< \widetilde{H}_{\rm rec} \big> = K^{(1)}$ does not depend on $M$. The
factor of $1/2$ removes the double counting coming from the presence of the recoil terms both in
the operator and in the wave function.

The quadratic recoil correction is known analytically for the hydrogen-like ions and numerically
for the helium atom \cite{pachucki:17:heSummary}. For a hydrogen-like system, the nonrelativistic
contribution to $K^{(2)}$ is induced by the reduced mass and is just opposite to the corresponding
contribution to $K^{(1)}$,
\begin{align}\label{eq:14}
K^{(2)}_{\rm nr}({\rm hydr}) = -K^{(1)}_{\rm nr}({\rm hydr})\,.
\end{align}
This can also be used as a reasonable approximation for the {\em nonrelativistic} contribution to
$K^{(2)}$ in few-electron systems (since the two-electron part of the nuclear recoil is usually
smaller than the one-electron part). However, the relativistic effects can cause significant
deviations from this simple formula. In particular, it was shown in
Ref.~\cite{pachucki:17:heSummary} that for helium the relativistic correction to  $K^{(2)}$ is much
larger that the corresponding correction to $K^{(1)}$. It is, therefore, not surprising that the
numerical calculations of $K^{(2)}$ performed in this work for the relativistic fine-structure
transitions show significant deviations from the simple nonrelativistic estimate (\ref{eq:14}).

\subsection{Other nonlinear effects}
\label{sec:nlinear}

It was pointed out in Ref.~\cite{flambaum:18} that for light atoms such as argon considered here,
the quadratic nuclear recoil is the main source of nonlinearity of the King plot within the
Standard Model framework. We now confirm this statement by examining other possible sources of a
nonlinearity.

First, we consider the energy correction that depends both on the nuclear mass $M$ and the nuclear
size $R$. The underlying physical effect is the fns correction to the nuclear recoil (``fns
recoil''). Strictly speaking, this is neither mass shift nor field shift, but we can formally
enforce the form of Eq.~(\ref{eq:1}) by ascribing the corresponding correction either to $K$ or to
$F$. It is tempting to try to calculate this effect numerically, e.g., by varying the nuclear
radius in the numerical code for the relativistic recoil correction. This would lead, however, to a
completely incorrect result. It was shown in Ref.~\cite{shabaev:98:recground} that the numerically
dominant fns recoil contribution coming from the relativistic operator (\ref{eq:rrec}) is spurious
and is exactly cancelled by the corresponding part of the QED fns recoil effect. Therefore, any
meaningful calculation of the fns recoil effect can be performed only within the framework of QED,
which is beyond the scope of the present paper.

In the present study, we estimate the fns recoil effect within the non-relativistic
independent-electron approximation. In this limit, the fns recoil effect is induced just by the
reduced-mass correction to the leading fns contribution, see, e.g., Ref.~\cite{mohr:16:codata}.
Therefore, within this approximation the leading field-shift constant should be multiplied by the
reduced-mass prefactor,
\begin{align}\label{eq:15}
F \to F \,\Big(1 - 3\, \frac{m}{M}\Big)\,.
\end{align}
We expect that this estimation gives the correct order of magnitude of the effect, even though we
are considering the fine-structure transitions, for which the nonrelativistic approximation does
not work well.

Another effect that may contribute to the nonlinearity of the King plot is the relativistic
correction to the field shift. For a light hydrogen-like atom, the numerically dominant
relativistic fns correction is delivered by the leading logarithmic approximation and is given by
(see, e.g., Ref.~\cite{mohr:16:codata})
\begin{align}\label{eq:16}
\delta E_{\rm fns} = \delta E_{\rm fns, nr}\Big[ 1 - (Z\alpha)^2\, \ln\Big(Z\alpha \frac{R}{\lambdabar_C}\Big) \Big]\,,
\end{align}
where $\delta E_{\rm fns, nr}$ is the nonrelativistic fns energy shift. Note that this relativistic
correction appears also in the fully relativistic approach \cite{shabaev:93:fns,flambaum:18},
originating through a modification of the exponent of the $R$ dependence of the fns energy shift,
\begin{align}\label{eq:17}
\Big(Z\alpha \frac{R}{\lambdabar_C}\Big)^2 \to \Big(Z\alpha \frac{R}{\lambdabar_C}\Big)^{2\gamma}\,,
\end{align}
where $\gamma = \sqrt{1-(\Za)^2}$.

Only the $R$-dependent part of the relativistic fns correction contributes to the nonlinearity of
the King plot. As an estimation, we assume it to have the same form as for the hydrogenic atoms,
\begin{align}\label{eq:18}
F \to F \,\bigg[ 1 - (Z\alpha)^2\, \ln\Big(\frac{R}{\lambdabar_C}\Big) \bigg]\,.
\end{align}

Finally, we consider the nuclear polarization, which is obviously isotope-dependent and thus
contributes to the nonlinearity of the King plot. Ref.~\cite{yerokhin:15:Hlike} reported the
following estimate for the nuclear-polarization energy shift $\delta E_{\rm npol}$, which is based
on available calculations for medium- and high-$Z$ ions,
\begin{align}\label{eq:19}
\delta E_{\rm npol} \approx -\frac1{1000}\,\delta E_{\rm fns} \pm 100\%\,.
\end{align}
This estimate gives a reasonable $Z$ scaling of the nuclear polarization but can significantly
underestimate the effect for the isotope shift. In order to correct for this, we introduce an
additional dependence on the mass number $A$,
\begin{align}\label{eq:19b}
\delta E_{\rm npol} \approx -\frac1{1000}\,\delta E_{\rm fns} \Big(\frac{A}{A_0}\Big)^n \pm 100\%\,,
\end{align}
where $A_0$ is the mass number of a selected isotope in the isotope chain and $n$ is an empirical
parameter. The giant resonance model of the nuclear polarizability by Migdal \cite{migdal:45} (see
also Ref.~\cite{flambaum:18}) yields $\delta E_{\rm npol} \propto R^2 A$, and thus $n = 1$.
Numerical calculations of the nuclear-polarization energy shifts for isotope chains of heavy H-like
ions \cite{plunien:95} suggest even larger values of $n$. For our estimates in the present work we
will use $n = 3$ and assume that it yields the expected order of magnitude of the effect. So, we
estimate the influence of the nuclear polarization on the nonlinearity of the King plot by applying
the following multiplicative factor to the field-shift constant,
\begin{align}\label{eq:20}
F \to F \,\bigg[ 1  - \frac1{1000}\,\Big(\frac{A}{A_0}\Big)^3 \bigg]\,.
\end{align}

\section{Calculations}

In the present work we investigate the $2P$ fine-structure transitions in Be-like, B-like, and
C-like argon, specifically, the $(1s)^22s2p\ ^3P_{2}\,$--$\,^3P_{1}$ transition in Ar$^{14+}$
(labeled as ``$a$''), the $(1s)^2(2s)^22p\ ^2P_{3/2}\,$--$\,^2P_{1/2}$ transition in Ar$^{13+}$
(labeled as ``$b$''), and the $(1s)^2(2s)^2(2p)^2\ ^3P_{1}\,$--$\,^3P_{0}$ transition in Ar$^{12+}$
(labeled as ``$c$''). We perform relativistic calculations of the mass-shift and field-shift
constants $K^{(1)}$, $K^{(2)}$, and $F$. The calculations are performed by the relativistic
configuration-interaction (CI) method with configuration-state wave functions (CSFs) constructed
with $B$-splines. Our implementation of the method is described in
Refs.~\cite{yerokhin:08:pra,yerokhin:12:lilike}.

The present calculations of the isotope-shift constants pose several difficulties. The first one
comes from the fact that we are considering the fine-structure transitions, for which all
relativistic effects are very much enhanced. The second one comes from strong mixing of the
reference states with the closely-lying levels. In order to take this into account, we perform the
CI expansions from multiple reference states, including the dominant mixing configurations as
additional reference states. In particular, we use the $1s^22s^2+1s^22p^2$ reference state for the
Be-like argon; the $1s^22s^22p+1s^22p^3$ reference state for the B-like argon and $1s^22s^22p^2 +
1s^2 2p^4$ for C-like argon. In the CI expansion we include the single, the double, and the
dominant part of triple and quadruple excitations from the multiple reference states specified
above. The CI expansions in this work includes up to 1.6~million CSFs.

Calculations of the quadratic mass-shift constant $K^{(2)}$ turn out to be significantly more
involved than those of the linear mass-shift constant $K^{(1)}$. The reason is that the relative
contribution of triple and quadruple (TQ) excitations are much more important for $K^{(2)}$ than
for $K^{(1)}$. In particular, for B-like Ar, the inclusion of the TQ excitations into the CI
expansion changes the result for $K^{(1)}$ on a 0.2-2\% level (depending on the choice of the
one-electron basis), whereas for $K^{(2)}$ it is a 30\% effect. For C-like Ar, the role of the TQ
excitations becomes even more significant: the inclusion of the TQ excitations reduces the result
for $K^{(2)}$ by an order of magnitude. The higher-order excitations are partly included through
the usage of multiple reference states, but a systematic study of such excitations is presently not
possible due to technical limitations for the size of the CSFs expansion. The numerical uncertainty
of the obtained results was estimated by varying the choice of the one-electron basis, which
changes the relative contributions of the individual excitations.

\section{Results and Discussion}

Numerical results of our relativistic calculations of the isotope-shift constants $K^{(1)}$,
$K^{(2)}$, and $F$ are presented in Table~\ref{tab:is}. The definition of the isotope-shift
constants is given by Eqs.~(\ref{eq:1}) and (\ref{eq:13}). We note that our results for $K^{(1)}$
do not include the QED part of the nuclear recoil, which was accounted for Be-like and B-like argon
in Refs.~\cite{orts:06,zubova:16}. The QED correction to $K^{(1)}$ does not cause a nonlinearity in
the King plot, so it is not considered in the present work. The results of our CI calculation for
the linear isotope-shift constants are in good agreement with previous relativistic calculations
\cite{orts:06,zubova:16,naze:14}.

Table~\ref{tab:Xphi} presents numerical results of our calculations of the ``new-physics''
isotope-shift constant $X_{\phi}$ defined by Eq.~(\ref{eq:7c}), for different values of masses of
the hypothetical boson $m_{\phi}$. Predictably, for small values of the boson mass $m_{\phi}$,
$\exp({-m_{\phi}r)/r} \approx 1/r$, so that $X_{\phi}$ does not depend on $m_{\phi}$.

Knowing the isotope-shift constants $K^{(1)}$, $K^{(2)}$, and $F$ for several transitions of the
same element, we can now calculate the modified isotope shifts $n_{xij}$ according to
Eq.~(\ref{eq:7}) and then the nonlinearity of the King plot according to Eq.~(\ref{eq:12b}). We
consider the King plots constructed for three pairs of transitions, $(a,b)$, $(b,c)$, and $(a,c)$,
and the chain of four isotopes of Ar with the mass numbers $(A_0,A_1,A_2,A_3) = (36, 38, 40, 42)$.

Using the values of the isotope-shift constants summarized in Table~\ref{tab:is}, we obtain the
following results for the King-plot nonlinearities caused by the quadratic recoil effect,
\begin{align}
\Delta_{\rm NL}(ab) &\ = 12.2\,(3)~{\rm kHz}\,,\\
\Delta_{\rm NL}(bc) &\ = 29.\,(7.)~{\rm kHz}\,,\\
\Delta_{\rm NL}(ac) &\ = 5.3\,(1.7)~{\rm kHz}\,.
\end{align}
We checked that other nonlinear effects discussed in Sec.~\ref{sec:nlinear} induce very small
contributions to $\Delta_{\rm NL}$. The largest of the subleading effects is the nuclear
polarization, whose contributions  to $\Delta_{\rm NL}$ for the transitions under consideration
were found to be $\sim 0.1\,$--$\,0.2$~kHz.

It is interesting that the King-plot nonlinearities calculated in this work are by 3-4 orders of
magnitude larger than the previous estimate (3~Hz) obtained for Ca$^+$ in Ref.~\cite{flambaum:18}.
The reason of such difference is not clear to us. It might be pointed out that no actual
calculations were performed in Ref.~\cite{flambaum:18}; only the expected order of magnitude of the
effect was estimated.

Our calculations demonstrate that isotope-shift measurements accurate at the Hertz level, like the
one reported for Ca$^+$ in Ref.~\cite{knollmann:19}, could no longer ignore nonlinearities
appearing in the Standard Model framework in the King-plot analysis. It is clear that such effects
should become observable in the near future. Specifically for argon isotopes, an experimental
identification of the nonlinear effects calculated in the present work is feasible by applying the
quantum-logic technique, as recently demonstrated for boron-like Ar$^{13+}$ \cite{micke:19:priv}.
With the same technique, the ground-state $(1s)^2(2s)^2(2p)^2\,\,^3\!P_{1}\,$--$\,^3\!P_{0}$
transition in carbon-like Ar$^{12+}$ at about 1015~nm could be resolved with a comparable level of
precision (of a few Hertz). The argon isotopes $^{36}$Ar, $^{38}$Ar, and $^{40}$Ar are stable and
affordable for such experimental studies. $^{42}$Ar is a $\beta$-emitter with a lifetime of 33
years and is in principle also accessible for this kind of experiment, arguably with some efforts
regarding procurement and safety requirements. More stable isotopes are available for calcium. The
same transitions in boron- and carbon-like Ca are still in the laser-accessible range for studying
the King-plot nonlinearities with even five isotopes, corresponding to four data points in the King
plot.

Recently, there was a suggestion put forward \cite{frigiuele:17,berengut:18} to use (the absence
of) the observed nonlinearity of the King plot in the isotope-shift measurements in order to
constrain the hypothetical new long-range forces between the electron and the nucleus.
Ref.~\cite{berengut:18} analyzed perspectives of such constraints for a (rather optimistic) variant
of the experimental accuracy of 1~Hz and the absence of King-plot nonlinearities on this level. Our
calculations show that the typical King-plot nonlinearities originating within the Standard Model
are much larger than 1~Hz; it is clear that we could constrain the new-physics effects only to the
level on which we are able to control the accuracy of the non-linear effects within the Standard
Model.

In order to predict which constraints on the new-physics coupling constant $\alpha_{\rm NP}$ in
Eq.~(\ref{eq:np:1}) one could expect from isotope-shift measurements of the transitions considered
in this work, we list in Table~\ref{tab:newphysics} the ratios $\alpha_{\rm NP}/\alpha$ that induce
a 1~kHz King-plot nonlinearity $\Delta_{\rm NL}$, for different masses of the hypotetical boson
$m_{\phi}$. We conclude that the perspective constraints are on a much more modest scale than was
anticipated in Ref.~\cite{berengut:18}. In order to obtain better constraints, one would need to
search for elements and/or transitions for which the Standard-Model King-plot nonlinearities are as
small as possible. In particular, investigations of heavier elements might be advantageous since
for them the nuclear recoil effects are suppressed due to a larger nuclear mass.

\section{Conclusion}

In this work we performed relativistic calculations of the isotope-shift constants for the $2P$
fine-structure transitions in Be-like, B-like, and C-like argon. For the first time, the quadratic
recoil constant $K^{(2)}$ in these systems was calculated. Because of significant contributions
from triple and quadruple excitations, large-scale configuration-interaction calculations with more
than a million configuration-state functions were employed, in order to obtain reliable predictions
for the quadratic mass-shift constant.

We studied nonlinear effects in the King plot for a chain of argon isotopes. It was demonstrated
that for such light atoms, the nonlinear effects in the King plot are dominated by the quadratic
recoil effect. For the considered fine-structure transitions, nonlinearities from 5 to 30~kHz were
found. Such effects should be clearly visible in the forthcoming isotope-shift experiments at the
Hertz accuracy level.

The nonlinear effects in the King plot arising within the Standard Model and the accuracy of their
theoretical description put limitations on possible constraints on hypothetical new long-range
forces between the electron and the nucleus, which can be derived from the isotope-shift
investigations. In the present work we performed calculations demonstrating to which level the
new-physics coupling constant can be realistically constrained for the considered transitions.

After the work reported in this paper have been finished, we learned about a new measurement of the
$^1S_0$--$^3P_{0,1}$ isotope shifts in strontium with a 10~kHz accuracy \cite{miyake:19}. The
authors observe a possible nonlinearity of the King plot and conclude that ``Future theoretical and
experimental studies should help to explain our observations...''.

\begin{acknowledgments}
Valuable conversations with I.~I.~Tupitsyn are gratefully acknowledged. The work is supported by
the German Research Foundation (DFG) under the Projects SU 658/4-1, SCHM2678/5-1 and the Cluster of
Excellence EXC2123 QuantumFrontiers. V.A.Y. acknowledges support by the Ministry of Education and
Science of the Russian Federation Grant No. 3.5397.2017/6.7.
\end{acknowledgments}

\begin{table*}
\caption{Relativistic isotope-shift constants for Be-like, B-like, and C-like argon, in a.u.
\label{tab:is}}
\begin{ruledtabular}
\begin{tabular}{lllddddd}
\multicolumn{1}{c}{Label} &
\multicolumn{1}{c}{Transition} &
\multicolumn{1}{c}{Ion}    &
\multicolumn{1}{c}{$K^{(1)}$}    &
\multicolumn{1}{c}{$K^{(2)}$}    &
\multicolumn{1}{c}{$F$}
\\
\hline\\[-5pt]
$a$ & $(1s)^22s2p\,\,^3\!P_{2}\,$--$\,^3\!P_{1}$                       & Ar$^{14+}$ & -0.1072\,(3)    &   0.289\,(3) & -0.000\,326\,(1)      \\
    &                                                                  &            & -0.107\,^b      &              & -0.000\,3\,^b    \\
    &                                                                  &            & -0.1072\,^c     &              & -0.000\,33\,^c   \\[5pt]
$b$ & $(1s)^2(2s)^22p\,\,^2\!P_{3/2}\,$--$\,^2\!P_{1/2}$               & Ar$^{13+}$ & -0.1900\,(3)    &  -0.202\,(35)& -0.001\,43\,(1)     \\
    &                                                                  &            & -0.1913\,^a     &              & -0.001\,4\,(1)^a \\
    &                                                                  &            & -0.1908\,^c     &              & -0.001\,45\,^c    \\[5pt]
$c$ & $(1s)^2(2s)^2(2p)^2\,\,^3\!P_{1}\,$--$\,^3\!P_{0}$
                                                                       & Ar$^{12+}$ & -0.0740\,(16)   &   0.310\,(68)& -0.000\,118\,(5) \\
    &                                                                  &            & -0.0735\,^c     &              & -0.000\,13\,^c   \\
\end{tabular}
\end{ruledtabular}
$^a$ CI-DFS \cite{zubova:16}, without QED,\\
$^b$ CI-DFS  \cite{orts:06}, without QED,\\
$^c$ MCDF \cite{naze:14}.
\end{table*}

\begin{table*}
\caption{``New-physics'' isotope-shift constant $X_{\phi}$, in a.u., for different values of
masses of the hypothetical boson $m_{\phi}$.
 \label{tab:Xphi}}
\begin{ruledtabular}
\begin{tabular}{cccccccc}
\multicolumn{1}{c}{Transition} &
\multicolumn{1}{c}{$m_{\phi}= 10$~eV} &
\multicolumn{1}{c}{$m_{\phi}= 10^2$~eV} &
\multicolumn{1}{c}{$m_{\phi}= 10^3$~eV} &
\multicolumn{1}{c}{$m_{\phi}= 10^4$~eV} &
\multicolumn{1}{c}{$m_{\phi}= 10^5$~eV} &
\multicolumn{1}{c}{$m_{\phi}= 10^6$~eV} &
\multicolumn{1}{c}{$m_{\phi}= 10^7$~eV}
\\
\hline\\[-5pt]
$a$ & $0.020$  & $0.020$  & $0.020$  & $0.018$ & $0.0029$ & $3.1\times10^{-5}$  & $2.8\times10^{-7}$\\
$b$ & $0.027$  & $0.027$  & $0.027$  & $0.024$ & $0.0050$ & $1.1\times10^{-4}$  & $1.3\times10^{-6}$\\
$c$ & $0.014$  & $0.014$  & $0.014$  & $0.013$ & $0.0019$ & $1.3\times10^{-5}$  & $1.\times10^{-7}$
\end{tabular}
\end{ruledtabular}
\end{table*}

\begin{table*}
\caption{Ratios of the ``new-physics'' coupling constant $\alpha_{\rm NP}$ to the
fine-structure constant $\alpha$ which would cause a nonlinearity of the King plot of 1~kHz, for different values of
masses of the hypothetical boson $m_{\phi}$.
 \label{tab:newphysics}}
\begin{ruledtabular}
\begin{tabular}{ccccccccc}
\multicolumn{1}{c}{} &
\multicolumn{1}{c}{Transitions} &
\multicolumn{1}{c}{$m_{\phi}= 10$~eV} &
\multicolumn{1}{c}{$m_{\phi}= 10^2$~eV} &
\multicolumn{1}{c}{$m_{\phi}= 10^3$~eV} &
\multicolumn{1}{c}{$m_{\phi}= 10^4$~eV} &
\multicolumn{1}{c}{$m_{\phi}= 10^5$~eV} &
\multicolumn{1}{c}{$m_{\phi}= 10^6$~eV} &
\multicolumn{1}{c}{$m_{\phi}= 10^7$~eV}
\\
\hline\\[-5pt]
$\nicefrac{\alpha_{\rm NP}}{\alpha}$
  & $(a,b)$
    & $1\times10^{-11}$  & $1\times10^{-11}$   & $1\times10^{-11}$   & $1\times10^{-11}$
       & $8\times10^{-11}$   & $2\times10^{-8}$  & $2\times10^{-5}$
       \\
  & $(b,c)$
    & $5\times10^{-12}$  & $5\times10^{-12}$   & $5\times10^{-12}$   & $6\times10^{-12}$
       & $4\times10^{-11}$   & $1\times10^{-8}$  & $6\times10^{-6}$
       \\
  & $(a,c)$
    & $1.5\times10^{-11}$  & $1.5\times10^{-11}$   & $1.5\times10^{-11}$   & $1.6\times10^{-11}$
       & $1\times10^{-10}$   & $2\times10^{-8}$  & $3\times10^{-5}$
\end{tabular}
\end{ruledtabular}
\end{table*}



\end{document}